\documentclass[reprint,aps,twoside,pre,showkeys,superscriptaddress,showpacs]{revtex4-1}
\usepackage{graphicx}
\usepackage{amsmath}
\usepackage{times}
\def\<{\left\langle}
\def\>{\right\rangle}
\def\onehalf{{\textstyle\frac{1}{2}}}
\def\quarter{{\textstyle\frac{1}{4}}}
\begin{document}
\title{Invariants for time-dependent Hamiltonian systems}
\author{J\"urgen Struckmeier and Claus Riedel}
\affiliation{Gesellschaft f\"ur Schwerionenforschung (GSI),
Planckstrasse~1, 64291~Darmstadt, Germany}
\date{Received 21 September 2000, published 17 July 2001}

\begin{abstract}
An exact invariant is derived for $n$-degree-of-freedom
Hamiltonian systems with general time-dependent potentials.
The invariant is worked out in two equivalent ways.
In the first approach, we define a special {\it Ansatz\/} for
the invariant and determine its time-dependent coefficients.
In the second approach, we perform a two-step canonical transformation
of the initially time-dependent Hamiltonian to a
time-independent one.
The invariant is found to contain a function of time $f_{2}(t)$,
defined as a solution of a linear third-order differential equation whose
coefficients depend in general on the explicitly known configuration space
trajectory that follows from the system's time evolution.
It is shown that the invariant can be interpreted as the time
integral of an energy balance equation.
Our result is applied to a one-dimensional, time-dependent, damped
non-linear oscillator, and to a three-dimensional system of Coulomb-interacting
particles that are confined in a time-dependent quadratic external potential.
We finally show that our results can be used to assess the
accuracy of numerical simulations of time-dependent Hamiltonian systems.

\bigskip\noindent
DOI: 10.1103/PhysRevE.64.026503
\end{abstract}
\maketitle
\section{Introduction}\label{sec:intro}
Analytical approaches to isolate conserved quantities
for a given physical system is a key objective
in the realm of Hamiltonian theory.
In the special case of an autonomous system, where the Hamiltonian
does not depend on time explicitly, one conserved quantity is
immediately found: the Hamiltonian itself that then represents
the system's total energy as a constant of motion.
Unfortunately, the Hamiltonians of most real physical
systems are explicitly time dependent, hence do not provide
directly a conserved quantity.

One of the first approaches to identify conserved quantities
for explicitly time-dependent systems has been worked out in the
context of the Lagrangian formalism by Noether~\cite{noether}.
Lutzky~\cite{lutzky} demonstrated that the well-known invariant
for the one-dimensional time-dependent harmonic
oscillator~\cite{cousny,lewis} follows straightforwardly
from Noether's theorem.
Subsequently, Chattopadhyay~\cite{chatto} extended this
work to derive invariants from this theorem for certain
one-dimensional non-linear systems.

Another approach to work out conserved quantities for explicitly
time-dependent Hamiltonian systems has been pursued by
Leach~\cite{leach}.
Performing a finite time-de\-pen\-dent canonical transformation,
he mapped the Hamiltonian of the time-dependent damped harmonic
oscillator onto a time-independent one.
Expressing this new Hamiltonian in terms of the old coordinates,
one immediately obtains an invariant in the original system.

A third way to find exact invariants for time-dependent classical
Hamiltonians has been worked out systematically by Lewis and
Leach\cite{lutzleach} using direct {\it Ans\"atze\/} with different
powers in the canonical momentum.

In this paper, we will show in Sec.~\ref{sec:ansatz}
and~\ref{sec:cantra} that both, the direct approach with an {\it Ansatz\/}
quadratic in the canonical momenta, as well as the canonical
transformation approach can straightforwardly be generalized to
$n$-degree-of-freedom Hamiltonian systems with general
time-dependent potentials.
In either case, the same invariant is obtained.
The invariant is found to contain an unknown function of time
$f_{2}(t)$, which is given as a solution of a linear third-order
differential equation, referred to as the auxiliary equation.
In general, this equation depends
on the system's spatial degrees of freedom.
As a consequence, the auxiliary equation can only be
integrated in conjunction with the equations of motion.

From the energy balance equation for time-dependent Hamiltonian systems,
it is shown that the invariant can be interpreted as the sum of the
system's time-varying energy content and the energy fed into or
detracted from it.

We will present two applications of our findings in Sec.~\ref{sec:appli}.
In the first example, the invariant and the associated auxiliary equation
is worked out for the one-dimensional system of the damped asymmetric spring.
It is shown that for the special case of a vanishing nonlinearity,
the invariant agrees with the harmonic oscillator result.
For the case of autonomous systems we will furthermore
demonstrate that a solution of the auxiliary equation with
$f_{2}(t)\ne\mbox{const}$ leads to a nontrivial invariant
that exists in addition to the invariant given by the Hamiltonian.

In the second example, we will examine the more challenging case of
a three-dimensional ensemble of $N$ Coulomb-interacting particles
of the same species that are confined within a time-dependent
quadratic external potential.
From the form of the related auxiliary equation,
it will become obvious that the function $f_{2}(t)$ represents
a kind of generalization of a beam envelope function.
It is shown that the function $f_{2}(t)$ may become unstable,
depending on the strength of the external focusing forces ---
similar to the behavior of envelope functions~\cite{struckrei}.

In Sec.~\ref{sec:acccheck} we will point out that
the existence of an invariant for explicitly
time-dependent Hamiltonian systems can be used to assess
the accuracy of numerical simulations of such systems.
In analogy to autonomous systems, where the actual conservation
of the Hamiltonian can be used as an accuracy criterion,
we may check in a simulation of an explicitly time-dependent
system to what extent the numerically obtained invariant differs from
the exact invariant of the ideal case.
\section{Ansatz approach}\label{sec:ansatz}
We consider an $n$-degree-of-freedom system of particles of the same
species moving in an explicitly time-dependent potential $V$ that
may be described by a Hamiltonian $H$ of the form
\begin{equation}\label{ham0}
H=\sum_{i=1}^{n}\frac{c(t)}{2}p_{i}^{2}+V\big(\{x\},t\big)\,.
\end{equation}
Herein, $c(t)$ is defined as an arbitrary twice differentiable
function of time that combines the particles' kinetic energy and a
velocity-dependent potential leading to isotropic friction forces
with linear velocity dependence.
For $c(t)\equiv 1$, the Hamiltonian (\ref{ham0}) thus describes
systems without friction~\cite{struckriedel}.
The curly braces denote the set of $n$ configuration
space variables $\{x\}=x_{1},\ldots,x_{n}$.

From the canonical equations, we derive for each degree of
freedom $i$ the equations of motion
\begin{equation}\label{speqm0}
\dot{x}_{i}=c(t)\,p_{i}\;,\qquad
\dot{p}_{i}=-\frac{\partial V(\{x\},t)}{\partial x_{i}}\,.
\end{equation}
With $\{p\}=p_{1},\ldots,p_{n}$ the set of canonical momenta,
a quantity
\begin{equation}\label{invar}
I=I\left(\{x\},\{p\},t\right)
\end{equation}
constitutes an invariant of the particle motion
if its total time derivative vanishes along the
phase-space path representing the system's time evolution
\begin{displaymath}
\frac{d\mbox{} I}{d t}=\frac{\partial I}{\partial t}+\sum_{i=1}^{n}\bigg[
\frac{\partial I}{\partial x_{i}}\dot{x}_{i}+
\frac{\partial I}{\partial p_{i}}\dot{p}_{i}\bigg]=0\,.
\end{displaymath}
We examine the existence of a conserved quantity (\ref{invar})
for a system described by Eq.~(\ref{ham0}) with a special
{\it Ansatz\/} for $I$ being at most quadratic in the momenta
\begin{equation}
I=\sum_{i}\Big[\onehalf f_{2}(t)\,p_{i}^{2}+f_{1}(x_{i},t)\,p_{i}\Big]+
f_{0}\big(\{x\},t\big)\,.\label{invar0}
\end{equation}
The set of functions $f_{2}(t)$, $f_{1}(x_{i},t)$, and
$f_{0}(\{x\},t)$ that render $I$ invariant are to be determined.
With the equations of motion (\ref{speqm0}), $dI/dt=0$ means explicitly
\begin{align}
\sum_{i}\bigg[&\frac{1}{2} p_{i}^{2}\frac{d\mbox{} f_{2}}{d t}+
p_{i}\frac{\partial f_{1}}{\partial t}+
p_{i}^{2}c\,\frac{\partial f_{1}}{\partial x_{i}}+
p_{i}c\,\frac{\partial f_{0}}{\partial x_{i}}\notag\\
&-\left(p_{i}f_{2}+
f_{1}\right)\frac{\partial V}{\partial x_{i}}\bigg]+
\frac{\partial f_{0}}{\partial t}=0\label{deritot}\,.
\end{align}
We now eliminate step by step the functions $f_{1}$ and $f_{0}$
contained in Eq.~(\ref{invar0}).
To this end, one may arrange the terms of Eq.~(\ref{deritot})
with regard to their powers in the momenta $p_{i}$.
Equation~(\ref{deritot}) is fulfilled if the coefficients pertaining
to the powers of the momenta vanish separately for each index $i$.
From the terms proportional to $p_{i}^{2}$, we thus get the condition
\begin{displaymath}
\onehalf\dot{f}_{2}(t) +
c(t)\,\frac{\partial f_{1}(x_{i},t)}{\partial x_{i}}=0\,.
\end{displaymath}
It follows that $f_{1}(x_{i},t)$ must be a linear function in $x_{i}$
\begin{equation}\label{f1}
f_{1}(x_{i},t)=-\frac{\dot{f}_{2}}{2c}\,x_{i}\,,
\end{equation}
omitting an integration constant that does not depend on the
configuration space variables.

For the terms linear in $p_{i}$, the condition derived from
Eq.~(\ref{deritot}) reads
\begin{equation}\label{linear}
\frac{\partial f_{1}}{\partial t}=
f_{2}(t)\,\frac{\partial V}{\partial x_{i}}-
c(t)\,\frac{\partial f_{0}}{\partial x_{i}}\,.
\end{equation}
On the other hand, $\partial f_{1}/\partial t$ is given as the
partial time derivative of Eq.~(\ref{f1})
\begin{equation}\label{df1dt}
\frac{\partial f_{1}}{\partial t}=\left(
\frac{\dot{f}_{2}\dot{c}}{2c^{2}}-\frac{\ddot{f}_{2}}{2c}\right)x_{i}\;.
\end{equation}
Inserting Eq.~(\ref{df1dt}) into Eq.~(\ref{linear}), and solving for the terms
containing the partial derivatives of the yet unknown but arbitrary
ancillary function $f_{0}(\{x\},t)$, one obtains
the following partial differential equation for $f_{0}$
\begin{equation}\label{tdf1}
\frac{\partial f_{0}}{\partial x_{i}}=\left(
\frac{\ddot{f}_{2}}{2c^{2}}-\frac{\dot{f}_{2}\dot{c}}{2c^{3}}\right)x_{i}+
\frac{f_{2}}{c}\,\frac{\partial V}{\partial x_{i}}\,.
\end{equation}
A function $f_{0}(\{x\},t)$ with partial derivative
(\ref{tdf1}) is obviously given by
\begin{equation}\label{f0}
f_{0}\big(\{x\},t\big)=
\Bigg(\frac{\ddot{f}_{2}}{c^{2}}-\frac{\dot{f}_{2}\dot{c}}{c^{3}}\Bigg)
\sum_{i}\quarter x_{i}^{2}+\frac{f_{2}}{c}\,V\big(\{x\},t\big)\,.
\end{equation}

The remaining terms of Eq.~(\ref{deritot}) do not depend on the
momenta $p_{i}$.
The third condition for $I$ to embody an invariant of the
particle motion thus writes, making use of Eq.~(\ref{f1})
\begin{equation}\label{rem}
\frac{\partial f_{0}}{\partial t}+\frac{\dot{f}_{2}}{2c}
\sum_{i}x_{i}\,\frac{\partial V}{\partial x_{i}}=0\,.
\end{equation}
In order to eliminate the {\it Ansatz\/} function $f_{0}$ contained
in Eq.~(\ref{rem}), we calculate the partial time derivative
of Eq.~(\ref{f0}), i.e., the time derivative at fixed $x_{i}$
\begin{align}
\frac{\partial f_{0}}{\partial t}=
&\left(\frac{\dddot{f_{2}}}{c^{2}}-\frac{3\ddot{f}_{2}\dot{c}}{c^{3}}-
\frac{\dot{f}_{2}\ddot{c}}{c^{3}}+\frac{3\dot{f}_{2}\dot{c}^{2}}
{c^{4}}\right)\sum_{i}\quarter x_{i}^{2}\notag\\
+&\left(\frac{\dot{f}_{2}}{c}-\frac{f_{2}\dot{c}}{c^{2}}\right) V+
\frac{f_{2}}{c}\,\frac{\partial V}{\partial t}\,.\label{df0dt}
\end{align}
Inserting Eq.~(\ref{df0dt}) into Eq.~(\ref{rem}), we finally
get a homogeneous linear third-order differential equation for
$f_{2}(t)$ that only depends on the configuration space variables
\begin{align}
\big(&\dot{f}_{2}\,c-f_{2}\,\dot{c}\big)\,V+
f_{2}\,c\,\frac{\partial V}{\partial t}+
\onehalf\dot{f}_{2}\,c\,\sum_{i}x_{i}\frac{\partial V}{\partial x_{i}}\notag\\
+\bigg(&\dddot{f_{2}}-\frac{3\ddot{f}_{2}\,\dot{c}+\dot{f}_{2}\,\ddot{c}}{c}+
\frac{3\dot{f}_{2}\,\dot{c}^{2}}{c^{2}}\bigg)
\sum_{i}\quarter x_{i}^{2}=0\,.\label{dgl1a}
\end{align}
The invariant $I$ is finally obtained if we insert Eqs.~(\ref{f1}),
(\ref{f0}), and the Hamiltonian (\ref{ham0}) into the
{\it Ansatz\/} (\ref{invar0})
\begin{equation}\label{invar1}
I=\frac{f_{2}}{c}H-\frac{\dot{f}_{2}}{2c}\sum_{i}x_{i}\,p_{i}+
\frac{\ddot{f}_{2}\,c-\dot{f}_{2}\,\dot{c}}{4c^{3}}\sum_{i}x_{i}^{2}\,.
\end{equation}
Reviewing our approach to work out the invariant (\ref{invar1}),
we recollect that equations of motion (\ref{speqm0}) have been
plugged into the expression for $dI/dt=0$ in Eq.~(\ref{deritot}).
This means that the subsequent Eq.~(\ref{dgl1a}) --- in conjunction
with the side condition $I=\text{const}$ from Eq.~(\ref{invar1}) ---
may be conceived as a conditional equation for a potential
$V(\{x\},t)$ that is consistent with a solution of the
equations of motion (\ref{speqm0}).

Vice versa, we may also assume the equations of motion
(\ref{speqm0}) to be {\em previously\/} solved.
Then, the trajectory $\{x(t)\}$, the potential $V(\{x(t)\},t)$,
and its partial derivatives constitute known coefficients of
Eq.~(\ref{dgl1a}) that depend on time only.
In this understanding, Eq.~(\ref{dgl1a}) embodies
an ordinary differential equation for $f_{2}(t)$.
The invariant (\ref{invar1}) then follows from the solution
path $(\{x(t)\},\{p(t)\})$ of Eqs.~(\ref{speqm0}), and from
$f_{2}(t)$ as a solution function of Eq.~(\ref{dgl1a}).
According to the existence and uniqueness theorem for linear
ordinary differential equations, a unique solution $f_{2}(t)$
of Eq.~(\ref{dgl1a}) exists --- and consequently the invariant $I$ ---
if $V$ and its partial derivatives are continuous along $\{x(t)\}$.

With $f_{2}(t)$ a solution of Eq.~(\ref{dgl1a}), we may directly show that
$dI/dt=0$ holds along solutions of the equations of motion (\ref{speqm0}).
Substituting Eqs.~(\ref{speqm0}) into the total time derivative of
Eq.~(\ref{invar1}), we find that the resulting equation agrees
with Eq.~(\ref{dgl1a}).
Hence, Eq.~(\ref{invar1}) provides a conserved quantity
as a time integral of Eq.~(\ref{dgl1a})
if and only if the system's evolution is governed by
the equations of motion (\ref{speqm0}).
We will use this relationship in Sec.~\ref{sec:acccheck} to
estimate the numerical error of computer simulations
of dynamical systems described by Eq.~(\ref{speqm0}).

Conversely, the invariant $I=\text{const}$ from Eq.~(\ref{invar1})
in conjunction with the third-order equation (\ref{dgl1a})
can easily be shown to imply the equations of motion (\ref{speqm0})
by inserting Eq.~(\ref{dgl1a}) into the total time derivative
of Eq.~(\ref{invar1}).
Since $dI/dt\equiv 0$ must hold for all solutions $f_{2}(t)$ of
Eq.~(\ref{dgl1a}), the respective sums of terms proportional to
$\ddot{f}_{2}(t)$, $\dot{f}_{2}(t)$, and $f_{2}(t)$ must vanish separately.
For the terms proportional to $\ddot{f}_{2}(t)$, this means
\begin{equation}\label{ident1}
\frac{\ddot{f}_{2}}{2c^{2}}\sum_{i}x_{i}
\big(\dot{x}_{i}-c(t)\,p_{i}\big)\equiv 0\,.
\end{equation}
The identity (\ref{ident1}) must be fulfilled for {\em all\/}
initial conditions $(\{x(0)\},\{p(0)\})$ and resulting phase-space
trajectories $(\{x(t)\},\{p(t)\})$ of the underlying dynamical system.
Consequently, the expression in parentheses must vanish
separately for each index $i$, thereby establishing the
first equation of motion (\ref{speqm0}).
For the remaining terms of $dI/dt\equiv 0$, we find
\begin{displaymath}
\sum_{i}
\left(f_{2}p_{i}-\frac{1}{2c(t)}\dot{f}_{2}x_{i}\right)
\left(\dot{p}_{i}+\frac{\partial V}{\partial x_{i}}\right)\equiv 0\,.
\end{displaymath}
Similar to the previous case, we may only fulfill this
equation in general for any solution $f_{2}(t)$ of
Eq.~(\ref{dgl1a}) and each index $i$ if the second
equation of motion (\ref{speqm0}) holds.

Summarizing, we may state that the triple made up by the
equations of motion (\ref{speqm0}), the third-order equation
(\ref{dgl1a}), and the invariant $I=\text{const}$ of
Eq.~(\ref{invar1}) forms a logical triangle: if two sides
are given at a time, the third can be deduced.

The physical interpretation of the invariant (\ref{invar1})
can be worked out considering the total time derivative of the
Hamiltonian (\ref{ham0}).
Making use of the canonical equations (\ref{speqm0}), we find
\begin{equation}\label{e-balance}
\frac{d}{dt}\left[\sum_{i=1}^{n}\onehalf c(t)\,p_{i}^{2}+V\right]-
\sum_{i=1}^{n}\onehalf\dot{c}(t)\,p_{i}^{2}-
\frac{\partial V}{\partial t}=0\,,
\end{equation}
which represents just the explicit form of the general theorem
$dH/dt=\partial H/\partial t$ for the Hamiltonian (\ref{ham0}).
Equation~(\ref{e-balance}) can be interpreted as an energy balance
relation, stating that the system's total energy change $dH/dt$
is quantified by the dissipation and the explicit
time dependence of the external potential.
Multiplying Eq.~(\ref{e-balance}) by the dimensionless quantity
$f_{2}/c$, and inserting $\partial V/\partial t$ according to the
auxiliary equation (\ref{dgl1a}), the resulting terms sum up
to the total time derivative
\begin{displaymath}
\frac{d}{dt}\left[\frac{f_{2}}{c}H-\frac{\dot{f}_{2}}{2c}
\sum_{i}x_{i}\,p_{i}+\frac{\ddot{f}_{2}\,c-\dot{f}_{2}\,
\dot{c}}{4c^{3}}\sum_{i}x_{i}^{2}\right]=0\,.
\end{displaymath}
The expression in brackets coincides with the invariant (\ref{invar1}).
With the initial conditions $f_{2}(0)/c(0)=1$,
$\dot{f}_{2}(0)\!=\!\ddot{f}_{2}(0)=0$ for the auxiliary equation
(\ref{dgl1a}), the invariant $I$ can now be interpreted as the
conserved {\em initial\/} energy $H_{0}$ for a nonautonomous
system described by the Hamiltonian (\ref{ham0}),
comprising both the system's time-varying energy content $H$
and the energy fed into or detracted from the system.

The meaning of $f_{2}(t)$ follows directly from the representation
(\ref{invar1}) of the invariant if the Hamiltonian $H$ is treated
formally as an independent variable: $I=I(\{x\},\{p\},t,H)$.
A vanishing total time derivative of the invariant $I$ then writes
\begin{align*}
\frac{dI}{dt}=&\left.\frac{\partial I}{\partial t}
\right|_{\{x\},\{p\},H}\!\!\!\!+\left.\frac{\partial H}{\partial t}
\frac{\partial I}{\partial H}\right|_{\{x\},\{p\},t}\\+
\sum_{i}\bigg(\dot{x}_{i}&\left.\frac{\partial I}{\partial x_{i}}
\right|_{\{p\},t,H}+
\dot{p}_{i}\left.\frac{\partial I}{\partial p_{i}}
\right|_{\{x\},t,H}\bigg)=0\,.
\end{align*}
Inserting $x_{i}$ and $p_{i}$ from the canonical equations
(\ref{speqm0}), and making again use of the auxiliary equation
(\ref{dgl1a}) to eliminate the third-order derivative
$\dddot{f_{2}}(t)$, we find the expected result
\begin{displaymath}
\left.\frac{\partial I}{\partial H}
\right|_{\{x\},\{p\},t}=\frac{f_{2}(t)}{c(t)}\,.
\end{displaymath}
$f_{2}/c$ thus provides the slope of the total energy $I$
with respect to the actual system energy $H$.

We finally note that for the special
case $\dot{c}\equiv\partial V/\partial t\equiv 0$, i.e., for autonomous
systems, $f_{2}(t)\equiv c\equiv 1$ is always solution of Eq.~(\ref{dgl1a}).
For this case, the invariant (\ref{invar1}) reduces to $I=H$,
hence provides the system's total energy, which is a known invariant
for Hamiltonian systems with no explicit time dependence.
Nevertheless, Eq.~(\ref{dgl1a}) also admits solutions
$f_{2}(t)\ne\mbox{const}$ for these systems.
We thereby obtain another nontrivial invariant
that exists in addition to the invariant representing
the energy conservation law.
This will be demonstrated in an example at the end of Sec.~\ref{sec:appli1}.
\section{Canonical transformation approach}\label{sec:cantra}
This approach aims to transform the Hamiltonian (\ref{ham0})
to a new Hamiltonian $\widetilde{H}$ that no longer
depends on time explicitly, hence embodies the total energy
of the transformed system as a constant of motion.
It happens that this procedure is most clearly performed in two steps.
In the first step, we canonically transform the Hamiltonian (\ref{ham0})
to a new set of coordinates $\{x\}\rightarrow\{\bar{x}\}$,
$\{p\}\rightarrow\{\bar{p}\}$ to obtain an intermediate Hamiltonian $\bar{H}$.
The explicitly time-dependent generating function of this transformation may
be expressed in terms of the new locations and the old momenta as
\begin{equation}\label{genfu}
F_{3}\big(\{\bar{x}\},\{p\},t\big)=\sum_{i=1}^{n}\left[\frac{\dot{f}_{2}(t)}
{4c(t)}\,\bar{x}_{i}^{2}-\sqrt{f_{2}(t)}\,\bar{x}_{i}p_{i}\right]\,.
\end{equation}
The coordinate transformation rules derived from (\ref{genfu}) are
\begin{align*}
x_{i} &=-\frac{\partial F_{3}}{\partial p_{i}} =\sqrt{f_{2}}\,\bar{x}_{i}\\
\bar{p}_{i} &= -\frac{\partial F_{3}}{\partial\bar{x}_{i}} =
\sqrt{f_{2}}\,p_{i}-\frac{\dot{f}_{2}}{2c}\bar{x}_{i}\,.
\end{align*}
In matrix notation, this phase-space preserving linear transformation writes
\begin{equation}\label{coortra}
\begin{pmatrix} x_{i}\\~\\ p_{i}\end{pmatrix}=
\begin{pmatrix}\sqrt{f_{2}} & 0\\~\\
\dot{f}_{2}\big/\big(2c\sqrt{f_{2}}\big) & 1\big/\sqrt{f_{2}}
\end{pmatrix}
\begin{pmatrix}\bar{x}_{i}\\~\\\bar{p}_{i}\end{pmatrix}\,.
\end{equation}
Expressed in the new (barred) coordinates, the partial time derivative
of the generating function (\ref{genfu}) follows as
\begin{equation}\label{dF3dt}
\frac{\partial F_{3}}{\partial t}=\sum_{i}\left[
\left(\frac{\ddot{f}_{2}}{c}-\frac{\dot{f}_{2}^{2}}{cf_{2}}-
\frac{\dot{f}_{2}\dot{c}}{c^{2}}\right)\quarter\bar{x}_{i}^{2}-
\frac{\dot{f}_{2}}{2f_{2}}\bar{x}_{i}\bar{p}_{i}\right]\,.
\end{equation}
Writing finally the old Hamiltonian $H$ in terms of the new coordinates,
the transformed Hamiltonian $\bar{H}=H+\partial F_{3}/\partial t$ is obtained as
\begin{equation}\label{ham1}
\bar{H}=\frac{c(t)}{f_{2}(t)}\left[\sum_{i=1}^{n}
\onehalf\bar{p}_{i}^{2}+\bar{V}\big(\{\bar{x}\},t\big)\right]
\end{equation}
with the potential $\bar{V}$ in the transformed system given by
\begin{align}
\bar{V}\big(\{\bar{x}\},t\big)=\Bigg(&\frac{f_{2}\ddot{f}_{2}}{c^{2}}-
\frac{\dot{f}_{2}^{2}}{2c^{2}}-\frac{f_{2}\dot{f}_{2}\dot{c}}{c^{3}}\Bigg)
\sum_{i=1}^{n}\quarter\bar{x}_{i}^{2}+\notag\\
&\frac{f_{2}}{c}\,V\Big(\left\{\sqrt{f_{2}}\,\bar{x}\right\},t\Big)\,.
\label{newpot}
\end{align}
The new potential $\bar{V}$ consists of two components, namely,
a term related to the original potential $V$, and an
additional quadratic potential that describes the linear
forces of inertia occurring due to the time-dependent linear
transformation (\ref{coortra}) to a new frame of reference.

Up to now, the function $f_{2}=f_{2}(t)$ contained in the
generating function (\ref{genfu}) has been defined as
an arbitrary regular function of time.
We now require the potential $\bar{V}(\{\bar{x}\},t)$ --- as defined
by Eq.~(\ref{newpot}) --- to be independent of time explicitly:
\begin{equation}\label{dgl1b}
\frac{\partial\bar{V}(\{\bar{x}\},t)}{\partial t}\stackrel{!}{=}0\,.
\end{equation}
This means that $f_{2}(t)$ is now tailored to eliminate the explicit
time dependence of Eq.~(\ref{newpot}) exactly at $\{\bar{x}\}$.
The explicit time dependence that is introduced if the original
potential $V(\{x\},t)$ is expressed in the new spatial coordinates yields
\begin{equation}\label{oldpot}
\frac{\partial V}{\partial\big(\sqrt{f}_{2}\,\bar{x}_{i}\big)}
\frac{\partial\big(\sqrt{f}_{2}\,\bar{x}_{i}\big)}{\partial t}=
\frac{\dot{f}_{2}}{2f_{2}}\;x_{i}\frac{\partial V}{\partial x_{i}}\,.
\end{equation}
In detail, Eq.~(\ref{dgl1b}) with (\ref{oldpot}) means
in terms of the old spatial coordinates
\begin{align}\label{dgl1c}
\frac{\partial\bar{V}}{\partial t}=
\bigg(&\frac{\dddot{f_{2}}}{c^{2}}-\frac{3\ddot{f}_{2}\,\dot{c}+
\dot{f}_{2}\,\ddot{c}}{c^{3}}+
\frac{3\dot{f}_{2}\,\dot{c}^{2}}{c^{4}}\bigg)
\sum_{i}\quarter x_{i}^{2}+\\
\bigg(&\frac{\dot{f}_{2}}{c}-\frac{f_{2}\,\dot{c}}{c^{2}}\bigg)\,V+
\frac{f_{2}}{c}\frac{\partial V}{\partial t}+
\frac{\dot{f}_{2}}{2c}\sum_{i}x_{i}
\frac{\partial V}{\partial x_{i}}=0\notag\,.
\end{align}
We observe that Eq.~(\ref{dgl1c}) agrees with the linear differential equation
(\ref{dgl1a}) for $f_{2}(t)$, as obtained in Sec.~\ref{sec:ansatz}.
Provided that $f_{2}$ is a solution of Eq.~(\ref{dgl1c}),
the explicit time dependence of transformed Hamiltonian $\bar{H}$
is imposed by the preceding factor $c/f_{2}$ only
\begin{equation}\label{ham1a}
\bar{H}=\frac{c(t)}{f_{2}(t)}\!\!\left[\sum_{i=1}^{n}
\onehalf\bar{p}_{i}^{2}+\bar{V}\big(\{\bar{x}\}\big)\right]\,.
\end{equation}
This explicit time dependence of the Hamiltonian (\ref{ham1a})
can be eliminated in the second step with the help of
a time-scale transformation $t\rightarrow\tau$ defined by
\begin{equation}\label{timescale}
\tau(t)=\int_{t_{0}}^{t}\frac{c(t')}{f_{2}(t')}\,dt'.
\end{equation}
With $\tau$ the independent variable, the canonical equations
\begin{displaymath}
\frac{d\bar{x}_{i}}{d\tau}=\frac{\partial\widetilde{H}}
{\partial\bar{p}_{i}}\,,\qquad
\frac{d\bar{p}_{i}}{d\tau}=-\frac{\partial\widetilde{H}}
{\partial\bar{x}_{i}}
\end{displaymath}
follow from Eq.~(\ref{ham1a}).
The new Hamiltonian $\widetilde{H}=\bar{H}\,f_{2}/c$ contained
herein no longer depends on time explicitly
\begin{equation}
\widetilde{H}=\sum_{i=1}^{n}\onehalf\bar{p}_{i}^{2}+
\bar{V}\big(\{\bar{x}\}\big)\,.
\label{ham2}
\end{equation}
Expressing (\ref{ham2}) in terms of the original coordinates according to
Eq.~(\ref{coortra}), we get an invariant $I$ of the original system $H$
\begin{align}
\widetilde{H}=&\frac{f_{2}}{c}\left[\sum_{i}
\frac{c}{2}\,p_{i}^{2}+V\big(\{x\},t\big)\right]\notag\\
&-\frac{\dot{f}_{2}}{2c}\sum_{i}x_{i}p_{i}+
\frac{\ddot{f}_{2}\,c-\dot{f}_{2}\,\dot{c}}{4c^{3}}
\sum_{i}x_{i}^{2}
\,=\,I\,,
\label{invar2}
\end{align}
which has been derived previously in Eq.~(\ref{invar1})
on the basis of the {\em Ansatz\/} (\ref{invar0}).
\section{Numerical examples}\label{sec:appli}
\subsection{Time-dependent damped asymmetric spring}\label{sec:appli1}
As a simple example, we investigate the one-dimensional non-linear
system of a time-dependent ``damped asymmetric spring.''
With $c(t)=\exp[-F(t)]$, its Hamiltonian is defined by
\begin{equation}\label{ham3}
H=\onehalf e^{-F(t)} p^{2}+
\big(\onehalf\omega^{2}(t)\,x^{2}+a(t)\,x^{3}\big)e^{F(t)}\,.
\end{equation}
Writing $f(t)\equiv\dot{F}(t)$, the equation of motion follows as
\begin{equation}\label{speq3}
\dot{x}=p\,e^{-F(t)}\,,\,\,
\ddot{x}+f(t)\,\dot{x}+\omega^{2}(t)\,x+3a(t)\,x^{2}=0\,.
\end{equation}
The invariant $I$ is immediately found writing
the general invariant (\ref{invar1}) for one degree of
freedom with the Hamiltonian $H$ given by Eq.~(\ref{ham3})
\begin{align}
I=\onehalf e^{2F(t)}\Big[
& f_{2}\dot{x}^{2}-\dot{f}_{2}x\dot{x}+
x^{2}\big\{\onehalf\ddot{f}_{2}+\onehalf\dot{f}_{2}f(t)\notag\\+
&f_{2}\omega^{2}(t)+2xf_{2}a(t)\big\}\Big]\,.\label{invar3}
\end{align}
The function $f_{2}(t)$ for this particular case is given
as a solution of the linear third-order ordinary differential equation
\begin{align}
\dddot{f_{2}}&+3\ddot{f}_{2}f(t)+\dot{f}_{2}\dot{f}(t)+
2\dot{f}_{2}f^{2}(t)\notag\\
&+4\dot{f}_{2}\omega^{2}(t)+4f_{2}f(t)\omega^{2}(t)+
4f_{2}\omega\dot{\omega}(t)\vphantom{\left(\dot{f}_{2}\right)}\notag\\
&+2x(t)\big[2f_{2}\dot{a}(t)+4f_{2}a(t)f(t)+5\dot{f}_{2}a(t)\big]=0\,,
\label{dgl3a}
\end{align}
which follows from Eq.~(\ref{dgl1a}) or, equivalently, from Eq.~(\ref{dgl1c}).
Since the particle trajectory $x=x(t)$ is explicitly contained in
Eq.~(\ref{dgl3a}), the solution $f_{2}(t)$ can only be obtained integrating
Eq.~(\ref{dgl3a}) simultaneously with the equation of motion (\ref{speq3}).

We may easily convince ourselves that $I$ is indeed a conserved quantity.
Calculating the total time derivative of Eq.~(\ref{invar3}),
and inserting the equation of motion (\ref{speq3}), we
end up with Eq.~(\ref{dgl3a}), which is fulfilled by definition
of $f_{2}(t)$ for the given trajectory $x=x(t)$.

The third-order equation (\ref{dgl3a}) may be converted into
a coupled set of first- and second-order equations.
The second-order equation
\begin{equation}\label{dgl3b}
\ddot{f}_{2}-\frac{\dot{f}_{2}^{2}}{2f_{2}}+\dot{f}_{2}f(t) +
2f_{2}\omega^{2}(t)=\frac{g_{x}(t)}{f_{2}}\,e^{-2F(t)}
\end{equation}
is equivalent to Eq.~(\ref{dgl3a}) if the time derivative of
$g_{x}(t)$, introduced in Eq.~(\ref{dgl3b}), is given by
\begin{equation}\label{dgl3c}
\dot{g}_{x}(t)=-2x(t)f_{2}\,e^{2F(t)}\big(2f_{2}\dot{a}+
4f_{2}af+5\dot{f}_{2}a\big)\,.
\end{equation}
With the help of the auxiliary equation (\ref{dgl3b}), the
invariant (\ref{invar3}) may be expressed in the alternative form
\begin{equation}
I=\frac{e^{2F(t)}}{2f_{2}}\big[{\big(f_{2}\dot{x}-
\onehalf\dot{f}_{2}x\big)}^{2}+2x^{3}f_{2}^{2}(t)\,a(t)\big]+
\frac{g_{x}(t)}{4f_{2}}x^{2}\,.\label{invar3a}
\end{equation}
In contrast to Eq.~(\ref{dgl3a}), the equivalent coupled set of equations
(\ref{dgl3b}) and (\ref{dgl3c}) does not contain the time derivatives
of the external functions $f(t)$ and $\omega(t)$.
The invariant (\ref{invar3a}) reduces to the well-known
invariant\cite{leach} for the time-dependent damped harmonic oscillator
if \mbox{$a(t)\equiv 0$}, which means that $g_{x}(t)=g_{0}={\rm const}$.
For this particular linear system, Eq.~(\ref{dgl3b})
no longer depends on the specific particle trajectory $x=x(t)$.
The solution functions $f_{2}(t)$ and $\dot{f}_{2}(t)$ then
apply to all trajectories emerging as integrals
of the equation of motion (\ref{speq3}) with $a(t)\equiv 0$.
With regard to the general form of the differential equation for $f_{2}(t)$,
as given by Eq.~(\ref{dgl1a}), we conclude that a decoupling from
the equations of motion (\ref{speqm0}) may occur for isotropic
linear systems only.

Another property of the linear system $[a(t)\equiv 0]$
follows directly from Eq.~(\ref{invar3a}).
For a positive integration constant $g_{x}(t)=g_{0}>0$,
one finds that $f_{2}(t)\,I\geq 0$.
Consequently, $f_{2}(t)$ can never change sign, thus remains non-negative
for the initial condition $f_{2}(0)>0$, which means that $I>0$.
The generating function (\ref{genfu}) then remains real at all times $t$,
and accordingly the Hamiltonian $\widetilde{H}$ of the transformed system.

On the other hand, $f_{2}(t)$ may change sign for the general
nonlinear system (\ref{speq3}), depending on the strength
of the nonlinear forces.
Then, the time-dependent canonical transformation (\ref{coortra})
becomes imaginary, which means that the autonomous system
ceases to exist as a physical system.
Under these circumstances, the particle motion within the
time-dependent nonlinear system can no longer be expressed as
the linearly transformed motion within a {\em real\/} autonomous system.

\begin{figure}[htb]
\centering
\includegraphics*[width=55mm,angle=-90]{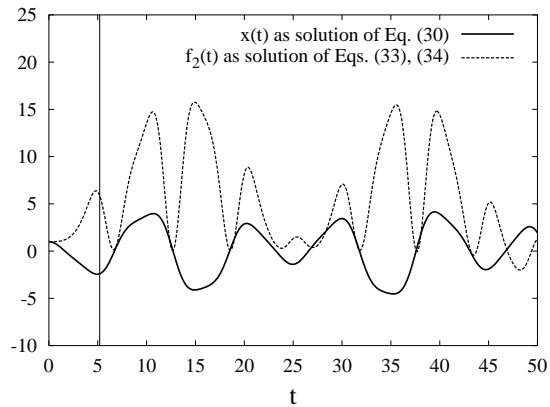}
\vspace*{3mm}
\caption{Example of a numerical integration of Eq.~(\ref{speq3})
and the simultaneous numerical integration of Eqs.~(\ref{dgl3b})
and (\ref{dgl3c}).
The vertical line marks the instant of time $t=5.2$, referred to
in Fig.~\ref{fig:ixp}.}
\label{fig:xoft}
\end{figure}

\begin{figure}[htb]
\centering
\includegraphics*[width=55mm,angle=-90]{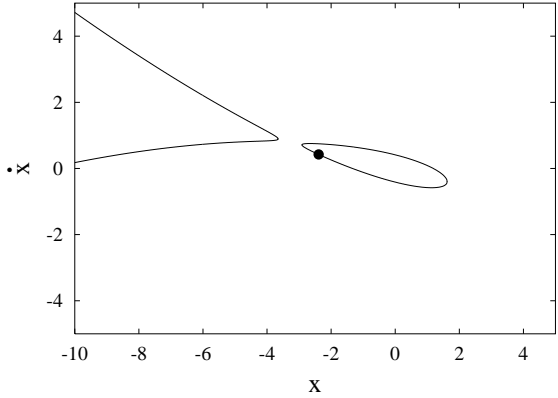}
\vspace*{1mm}
\caption{Lines of constant invariant $I=0.5$ in the
$(x,\dot{x})$ phase-space plane and location of the
sample particle at $t=5.2$, $f_{2}=6.1$.}
\label{fig:ixp}
\end{figure}

Figure~\ref{fig:xoft} shows a special case of a numerical integration
of the equation of motion (\ref{speq3}).
Included in this figure, we see the result of a simultaneous
numerical integration of Eqs.~(\ref{dgl3b}) and (\ref{dgl3c}).
The coefficients of Eq.~(\ref{speq3}) are defined as
$\omega(t)=\cos(t/2)$, $a(t)=5\times 10^{-3}\sin(t/3)$, and
\mbox{$f(t)=1.76\times 10^{-3}\cos^{2}(t/\pi)$}.
The initial conditions were set to $x(0)=1$, $\dot{x}(0)=0$,
$f_{2}(0)=1$, $\dot{f}_{2}(0)=0$, and $\ddot{f}_{2}(0)=0$.
According to Eq.~(\ref{invar3a}), we hereby define an invariant of
$I=0.5$ for the sample particle.
With both results, we are able to calculate the phase-space curve
of constant invariant $I(x,\dot{x},t)=0.5$, as defined by Eq.~(\ref{invar3a}).
\vspace*{-6mm}
Figure~\ref{fig:ixp} displays both a snapshot of this curve at $t=5.2$,
and the instantaneous location of the sample particle.
As expected, the particle lies exactly on the line of constant $I$,
thereby providing a numerical verification of Eq.~(\ref{invar3a}).
\vspace*{-6mm}
For the special case of an autonomous system, we define $c(t)=\exp(-F)=1$,
$\omega(t)=\omega_{0}=\mbox{const}$, and $a(t)=a_{0}=\mbox{const}$.
It follows that $\dot{F}=f=\dot{f}=\dot{\omega}=\dot{a}=0$,
which means that Eq.~(\ref{dgl3a}) reduces to
\begin{equation}\label{dgl3d}
\dddot{f_{2}}+\dot{f}_{2}\left(4\omega_{0}^{2}+10x(t)\,a_{0}\right)=0\,.
\end{equation}
Obviously, this equation has the special solution $f_{2}(t)\equiv 1$.
In that case, the invariant (\ref{invar3}) is given by
\begin{displaymath}
I=\onehalf\dot{x}^{2}+\onehalf\omega_{0}^{2}x^{2} + a_{0}x^{3} = H\,,
\end{displaymath}
thus coincides with the system's Hamiltonian that
represents the conserved total energy.
A further nontrivial invariant is obtained for solutions of
Eq.~(\ref{dgl3d}) with $f_{2}(t)\ne\mbox{const}$ as
\begin{equation}\label{invar3b}
I=\onehalf\big[ f_{2}\dot{x}^{2}-\dot{f}_{2}x\dot{x}+
x^{2}\big\{\onehalf\ddot{f}_{2}+
f_{2}\omega_{0}^{2}+2xf_{2}a_{0}\big\}\big]\,.
\end{equation}
For the harmonic oscillator, i.e., for $a_{0}=0$,
we may substitute the integral of Eq.~(\ref{dgl3d})
\begin{equation}\label{dgl3e}
\ddot{f}_{2}(t)+4\omega_{0}^{2}\,f_{2}(t)=0
\end{equation}
into Eq.~(\ref{invar3b}) to derive the invariant in the form
\begin{displaymath}
I=\onehalf\big[ f_{2}\big(\dot{x}^{2}-\omega_{0}^{2}x^{2}\big)
-\dot{f}_{2}x\dot{x}\big]\,.
\end{displaymath}
With $f_{2}(t)$ a solution of Eq.~(\ref{dgl3e}), this expression agrees
with the invariant presented earlier by Lutzky~\cite{lutzky2}.
\subsection{System of Coulomb-interacting particles}
We now analyze a three-dimensional example, namely, an ensemble of $N$
Coulomb-interacting particles of the same species moving in a
time-dependent quadratic external potential,
as typically given in the co-moving frame for charged particle beams
that propagate through linear focusing lattices.
The particle coordinates in the three spatial directions are
distinguished by $x_{i}$, $y_{i}$, and $z_{i}$, the canonical
momenta correspondingly by $p_{x,i}$, $p_{y,i}$, and $p_{z,i}$.
Setting $c(t)\equiv 1$ in Eq.~(\ref{ham0}), the Hamiltonian $H$
of this system may now be written as
\begin{equation}\label{ham4}
H =\sum_{i=1}^{N}\onehalf\big(p_{x,i}^{2}+p_{y,i}^{2}+
p_{z,i}^{2}\big)+V\big(\{x\},\{y\},\{z\},t\big)\,.
\end{equation}
The effective potential contained herein is given by
\begin{align}
V\big(\{x\},\{y\},\{z\},t\big)=\sum_{i=1}^{N}\Big[
&\onehalf\omega_{x}^{2}(t)\,x_{i}^{2}+
\onehalf\omega_{y}^{2}(t)\,y_{i}^{2}\notag\\+
&\onehalf\omega_{z}^{2}(t)\,z_{i}^{2}+
\onehalf\sum_{j\ne i}\frac{c_{1}}{r_{i j}}\Big]\,,\label{effpot}
\end{align}
with
$r_{i j}^{2}={(x_{i}-x_{j})}^{2}+{(y_{i}-y_{j})}^{2}+{(z_{i}-z_{j})}^{2}$
and \mbox{$c_{1}=q^{2}/4\pi\epsilon_{0}m$},
$q$ and $m$ denoting the particles' charge and mass, respectively.
The equations of motion that follow from Eq.~(\ref{speqm0})
with Eq.~(\ref{effpot}) are
\begin{equation}\label{speqm}
\dot{x}_{i} = p_{x,i}\,,\quad\ddot{x}_{i}+\omega_{x}^{2}(t)
\,x_{i}-c_{1}\sum_{j\ne i}\frac{x_{i}-x_{j}}{r_{i j}^{3}}=0\,,
\end{equation}
and likewise for the $y$ and $z$ directions.
We note that the factor $1/2$ in front of the Coulomb interaction term
in Eq.~(\ref{effpot}) disappears in Eq.~(\ref{speqm}) since each term
occurs twice in the symmetric form of the double sum.

For the effective potential (\ref{effpot}) and $c(t)\equiv 1$, the
third-order differential equation (\ref{dgl1a}) for $f_{2}$ specializes to
\begin{align}
\!\sum_{i}\Big[&x_{i}^{2}\big(\dddot{f_{2}}\!+\!4\dot{f}_{2}\omega_{x}^{2}\!+\!
4f_{2}\omega_{x}\dot{\omega}_{x}\big)\!+\!
y_{i}^{2}\big(\dddot{f_{2}}\!+\!4\dot{f}_{2}\omega_{y}^{2}\!+\!
4f_{2}\omega_{y}\dot{\omega}_{y}\big)\notag\\
+&\,z_{i}^{2}\big(\dddot{f_{2}}\!+\!4\dot{f}_{2}\omega_{z}^{2}\!+\!
4f_{2}\omega_{z}\dot{\omega}_{z}\big)\!+\!
\dot{f}_{2}\sum_{j\ne i}\frac{c_{1}}{r_{i j}}\Big]=0
\label{dgl2a}\,.
\end{align}
With $f_{2}(t)$ a solution of Eq.~(\ref{dgl2a}) and $H$ the Hamiltonian
(\ref{ham4}), the invariant follows directly from Eq.~(\ref{invar1}) as
\begin{align}
I=f_{2}(t)\,H&-\onehalf\dot{f}_{2}\sum_{i}
\big(x_{i}\,p_{x,i}+y_{i}\,p_{y,i}+z_{i}\,p_{z,i}\big)\notag\\
&+\quarter\ddot{f}_{2}\sum_{i}\big(x_{i}^{2}+y_{i}^{2}+
z_{i}^{2}\big)\,.\label{invar1a}
\end{align}
Equation~(\ref{dgl2a}) may be cast into a compact form if the sums over the
particle coordinates are written in terms of ``second beam moments,''
denoted as $\<x^{2}\>$ for the $x$ coordinates.
Likewise, the double sum over the Coulomb interaction terms may be
expressed as electric field energy $W(t)$ of all particles
\begin{displaymath}
\<x^{2}\>\!(t)=\frac{1}{N}\sum_{i}x_{i}^{2}(t)\,,\quad
W(t)=\frac{m}{2}\sum_{i}\sum_{j\ne i}\frac{c_{1}}{r_{i j}}\,.
\end{displaymath}
A similar notation will be used for all quadratic
terms of the particle coordinates.
Corresponding to the previous example, the third-order equation
(\ref{dgl2a}) may be split into a coupled set of first- and
second-order differential equations.
Similar to Eq.~(\ref{dgl3b}), we define the function $g=g(t)$ by
\begin{equation}\label{dgl4a}
f_{2}\ddot{f}_{2}-\onehalf\dot{f}_{2}^{2}+
2f_{2}^{2}\,\omega^{2}(t)=g(t)\,.
\end{equation}
The function $\omega^{2}(t)$ contained herein is defined
as the ``average focusing function'' according to
\begin{displaymath}
\omega^{2}(t)=\frac{\omega_{x}^{2}\<x^{2}\>+
\omega_{y}^{2}\<y^{2}\>+\omega_{z}^{2}\<z^{2}\>}
{\<x^{2}\>+\<y^{2}\>+\<z^{2}\>}\,.
\end{displaymath}
Comparing the time derivative of Eq.~(\ref{dgl4a}) with Eq.~(\ref{dgl2a}),
one finds that the time derivative of $g(t)$ must satisfy
\begin{gather}
\dot{g}(t)=\frac{1}{\<x^{2}\>+\<y^{2}\>+\<z^{2}\>}\bigg[
-2f_{2}\dot{f}_{2}\frac{W}{mN}+4f_{2}^{2}\big\{
\notag\\
\<xp_{x}\>\left(\omega_{x}^{2}-\omega^{2}\right)\!+\!
\<yp_{y}\>\left(\omega_{y}^{2}-\omega^{2}\right)\!+\!
\<zp_{z}\>\left(\omega_{z}^{2}-\omega^{2}\right)\!\big\}\bigg]\,.
\label{dgl4b}
\end{gather}
Unlike the third-order equation (\ref{dgl2a}), the equivalent
coupled set of equations (\ref{dgl4a}) and (\ref{dgl4b})
no longer contains the time derivatives of the external focusing
functions $\omega_{x}(t)$, $\omega_{y}(t)$, and $\omega_{z}(t)$.
We observe that $\dot{g}(t)$ is determined by two quantities
of different physical nature: the field energy constituted by all
particles as a measure for the strength of the Coulomb interaction,
and the system's anisotropy.
In contrast to $g_{x}(t)$ of the one-dimensional example of
Sec.~\ref{sec:appli1}, the function $g(t)$ is generally {\em not\/}
constant in the linear case, which is given here for a vanishing
Coulomb interaction ($W\rightarrow 0$).

With the help of Eq.~(\ref{dgl4a}), we may substitute $\ddot{f}_{2}(t)$
and the external focusing functions in Eq.~(\ref{invar1a}) to express
the invariant in the alternative form
\begin{align}
2f_{2}I/N =&\<{\big(f_{2}p_{x}-\onehalf\dot{f}_{2}x\big)}^{2}\>+
\<{\big(f_{2}p_{y}-\onehalf\dot{f}_{2}y\big)}^{2}\>\notag\\\mbox{}
+&\<{\big(f_{2}p_{z}-\onehalf\dot{f}_{2}z\big)}^{2}\>+
f_{2}^{2}(t)\frac{2W}{mN}\notag\\\mbox{}
+&\onehalf g(t)\left(\<x^{2}\>+\<y^{2}\>+\<z^{2}\>\right)\,.
\label{invar1b}
\end{align}
Similar to the previous example, the function $g(t)$
accounts for an eventual change of sign of $f_{2}(t)$,
owing to the fact that all other terms on the right hand
side of Eq.~(\ref{invar1b}) may not turn negative.

The canonical transformation (\ref{coortra}) becomes undefined
for instants of time $t$ with $f_{2}(t)=0$.
Furthermore, for time intervals with a negative value of $f_{2}(t)$,
the elements of the transformation matrix (\ref{coortra}) turn imaginary.
For these cases, the equivalent autonomous system of Eqs.~(\ref{ham4})
and (\ref{effpot}) that is defined by the canonical transformation
rules (\ref{coortra}) and (\ref{timescale})
ceases to exist in a physical sense.
This indicates that the beam evolves within the nonautonomous
system in a way that can no longer be correlated to the beam
evolution within an autonomous system by the linear canonical
transformation (\ref{coortra}).
In contrast, the invariant (\ref{invar1a}) itself exists for
{\em all\/} $f_{2}(t)$ that are solutions of the auxiliary
equation (\ref{dgl2a}).

\begin{figure}[htb]
\centering
\includegraphics*[width=57mm,angle=-90]{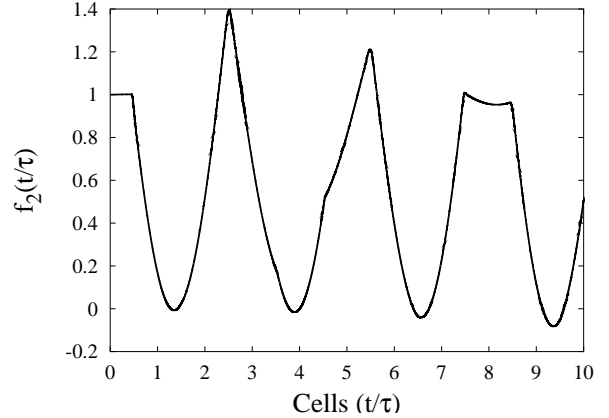}
\vspace*{2mm}
\caption{$f_{2}(t)$ as stable solution of Eq.~(\ref{dgl2a})
for $\sigma_{0}=45^{\circ}$, $\sigma=9^{\circ}$.
$\tau$~denotes the focusing period common to all three directions.}
\label{fig:f2s}
\end{figure}

\begin{figure}[htb]
\centering
\includegraphics*[width=57mm,angle=-90]{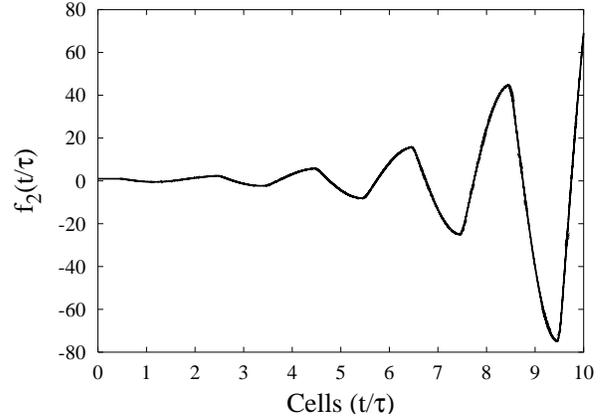}
\vspace*{2mm}
\caption{$f_{2}(t)$ as unstable solution of Eq.~(\ref{dgl2a})
for $\sigma_{0}=60^{\circ}$, $\sigma=15^{\circ}$.
$\tau$~denotes the focusing period common to all three directions.}
\label{fig:f2i}
\end{figure}

Figures~\ref{fig:f2s} and~\ref{fig:f2i} show the function $f_{2}(t)$
as the result of numerical integrations of the coupled set
(\ref{dgl4a}) and (\ref{dgl4b}).
The second-order moments --- denoted by the angle brackets ---
and the field energy function $W(t)$ were taken from simulations of a
fictitious three-dimensional anisotropic focusing lattice that is
described by the Hamiltonian (\ref{ham4}) with the potential (\ref{effpot}).
The simulation leading to Fig.~\ref{fig:f2s} was performed at
the zero-current tune of $\sigma_{0}=45^{\circ}$, and a space charge
depressed tune of $\sigma=9^{\circ}$ in each direction.
\vspace*{-8mm}
As a result of various simulations, we found that $f_{2}(t)$
becomes unstable for $\sigma_{0}\ge 60^{\circ}$.
Furthermore, it turned out that this limit value for an unstable
evolution of $f_{2}(t)$ decreases as the field energy $W(t)$ increases.
A case with a growing amplitude of $f_{2}(t)$ is displayed in
Fig.~\ref{fig:f2i} for a beam propagating under the conditions
of a zero-current tune of $\sigma_{0}=60^{\circ}$
and the depressed tune of $\sigma=15^{\circ}$.
In agreement with earlier studies on high current beam
transport~\cite{hola}, the simulation results show that
the beam moments remain bounded under these conditions.
This means that an instability of $f_{2}(t)$ is {\em not\/}
necessarily associated with an instability of the beam moments.
Nevertheless, the phase-space planes of constant $I$ become more
and more distorted as $f_{2}(t)$ and its derivatives diverge.
This may indicate a transition from a regular to a chaotic motion
of the beam particles.
\vspace*{-8mm}
\section{Checking the accuracy of numerical
simulations of Hamiltonian systems}\label{sec:acccheck}
The conserved quantity $I$ that has been shown to exist for
explicitly time-dependent Hamiltonian systems can be used
to test the results of numerical simulations of such systems.
As already stated in Sec.~\ref{sec:ansatz}, Eq.~(\ref{invar1})
embodies a time integral of Eq.~(\ref{dgl1a}) if the system's time
evolution is {\em strictly\/} consistent with the equations of
motion (\ref{speqm0}).
In the ideal case, i.e., if no numerical inaccuracies were
included in a computer simulation of a system governed by Eq.~(\ref{ham0}),
and no numerical errors were added performing the subsequent
integration of Eq.~(\ref{dgl1a}), we would not see any
deviation $\Delta I/I_{0}$ calculating the invariant
(\ref{invar1}) as a function of time.

Of course, we can never avoid numerical errors in
computer simulations of dynamical systems because of the
generally limited accuracy of numerical methods.
For the same reason, the numerical integration of Eq.~(\ref{dgl1a})
is also associated with a specific finite error tolerance.
Under these circumstances, the quantity $I$ as given by
Eq.~(\ref{invar1}) --- with $f_{2}(t)$, $\dot{f}_{2}(t)$,
and $\ddot{f}_{2}(t)$ following from Eq.~(\ref{dgl1a}) ---
can no longer be expected to be {\em exactly\/} constant.
Both numerical tasks --- the numerical integration of the
equations of motion (\ref{speqm0}), and the subsequent numerical
integration of Eq.~(\ref{dgl1a}) contribute to a nonvanishing
$\Delta I/I_{0}$ along the integration time span.
Nevertheless, since both tasks do not depend on each other
with respect to their specific error tolerances, we can regard
the obtained $\Delta I/I_{0}$ curve as a cross check of both
numerical methods.
Since the error tolerance for the numerical integration of
Eq.~(\ref{dgl1a}) is a known property of the underlying
algorithm, we can estimate from $\Delta I(t)/I_{0}$
the error tolerance integrating the equations of motion (\ref{speqm0}).

\vspace*{-6mm}
\begin{figure}[htb]
\centering
\includegraphics*[width=57mm,angle=-90]{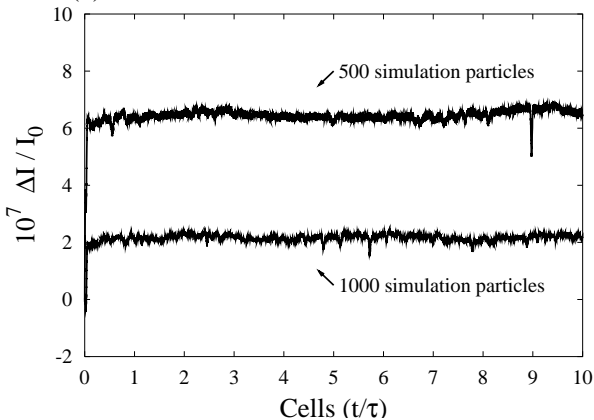}
\vspace*{3mm}
\caption{Relative invariant error $\Delta I/I_{0}$ for
three-dimensional simulations of a charged particle beam
with different numbers of macro-particles.}
\label{fig:inerr}
\end{figure}
Figure~\ref{fig:inerr} displays two examples of curves of relative
deviations $\Delta I/I_{0}$ from the invariant (\ref{invar1a})
for numerical simulations of a charged particle beam.
The function $f_{2}(t)$ and its derivatives that were used to calculate
$I$ were obtained from a numerical integration of Eq.~(\ref{dgl2a})
--- or equivalently from the coupled set (\ref{dgl4a}) and (\ref{dgl4b}).
The time-dependent coefficients of Eq.~(\ref{dgl2a}), namely, the second beam
moments and the field energy $W(t)$, had been determined
before from three-dimensional simulations
of charged particle beams propagating through a linear
focusing lattice with non-negligible Coulomb interaction,
as described by the potential function (\ref{effpot}).
As expected, the residual deviation $\Delta I/I_{0}$
depends on the number of macroparticles used in the simulation.

\begin{figure}[htb]
\centering
\includegraphics*[width=57mm,angle=-90]{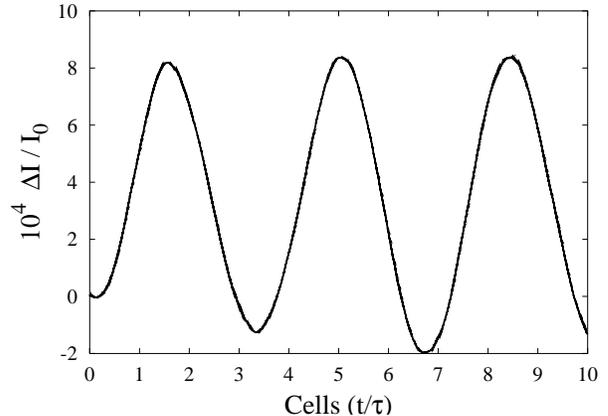}
\vspace*{3mm}
\caption{Relative invariant error $\Delta I/I_{0}$ for a
three-dimensional simulation of a charged particle beam with
$5\%$ error in the space charge force calculations.}
\label{fig:ierr}
\end{figure}

For a comparison, the corresponding deviation is plotted in
Fig.~\ref{fig:ierr} for a simulation with a systematic
$5\%$ error in the space charge force calculations.
We now find a relative deviation $\Delta I/I_{0}$
in the order of $10^{-3}$, hence three orders of magnitude larger
than the previous case with no artificial space charge force error.
\vspace*{-6mm}
By comparing simulation runs with different parameters, such as
the number of macroparticles, the time step size, and details of
the numerical algorithm used to integrate the equations of motion,
we may straightforwardly check whether the overall accuracy
of our particular simulation has been improved.
\section{Conclusions}
A fairly general result has been found: a conserved quantity
can straightforwardly be deduced for explicitly time-dependent
Hamiltonian systems.
The invariant contains an unknown function $f_{2}(t)$ and its first
and second time derivatives, which is determined by a linear
homogeneous third-order auxiliary differential equation.
In general, this auxiliary equation depends on the
system's spatial degrees of freedom.
Under these circumstances, the solution $f_{2}(t)$
can only be determined integrating the auxiliary equation
{\em simultaneously\/} with the equations of motion.
The invariant can be regarded as the conserved global energy
for nonautonomous systems, which is obtained if we add
to the time-varying energy represented by the Hamiltonian $H$
the energies fed into or detracted from the system.

The invariant has been found to agree with the known
conserved quantity of the one-dimensional time-dependent
harmonic oscillator~\cite{lewis,leach}.
For this particular one-dimensional linear case, the dependence
of the auxiliary equation on the particle position cancels.
Then the third-order auxiliary equation can directly be
integrated to yield a nonlinear second-order equation
for $f_{2}(t)$ that applies to all particle trajectories.
Furthermore, the second invariant for the time-independent harmonic
oscillator could straightforwardly be reproduced~\cite{lutzky2}.
All these invariants follow as special cases from the general
expressions of our invariant and the associated auxiliary equation.

The existence of an invariant has been shown to be
useful to check the accuracy of numerical simulations
of explicitly time-dependent Hamiltonian systems.
Having numerically integrated the equations of motion,
the system's third-order auxiliary differential equation can be
integrated, and the numerical value of the ``invariant'' $I$
can be calculated subsequently.
The relative deviation $\Delta I/I_{0}$ of $I$ from the exact
invariant $I_{0}$ can then be used as a measure for the accuracy
of the respective simulation.

The physical implications that are associated with an unstable
behavior of $f_{2}(t)$ of the auxiliary
differential equation remain to be investigated.
Furthermore, the physical meaning of solutions of the auxiliary
equation with $f_{2}(t)$ turning negative must be clarified.
In that case, the elements of coordinate transformation matrix
(\ref{coortra}) become imaginary, which means that the equivalent
autonomous system ceases to exist as a physical system.
This indicates that the explicitly time-dependent Hamiltonian
system evolves in a way that can no longer be correlated to
the evolution of a time-independent system by a linear mapping.
Nevertheless, the invariant $I$ of the explicitly time-dependent system
exists independently of the sign of $f_{2}(t)$.

We finally note that the invariant (\ref{invar1}) together with the
related auxiliary equation (\ref{dgl1a}) can be derived equivalently
performing an infinitesimal canonical transformation in the
extended phase-space.
Furthermore, the invariant and the auxiliary equation may be worked
out as well on the basis of Noether's theorem~\cite{noether}.
Our invariant thus embodies exactly the conserved quantity that
emerges as the result of Noether's symmetry transformation.
We will report these results in a forthcoming paper~\cite{annalen}.

\end{document}